\documentstyle[aps,prl,multicol]{revtex}

\def\date#1{\author{\small(#1)}}
\def\abstract#1{\author{\parbox[t]{5.5in}{\small#1}}\par\maketitle}

\def\f.#1.{{\bf #1}}
\def\mb.#1.{\bbox{#1}}

\def\Sec#1{\section{#1}} 
\def\opn{\begin{equation}} \def\cls{\end{equation}}
\def\opa{\begin{eqnarray}} \def\cla{\end{eqnarray}}
\def\opbib{\begin{references}} \def\clbib{\end{references}}
\def\bb#1{\bibitem{#1}}
\def\qno;#1;{\label{#1}\end{equation}} \def\qna;#1;{\label{#1}\end{eqnarray}}
\def\rf;#1;{(\ref{#1})}
\def\dels#1 {\nabla\kern -1.5pt_{#1}\kern 1.5pt}
 \def\av#1{\left\langle #1\right\rangle}
\def\avv#1{\langle #1\rangle}

\def\suspend{\end{multicols}\vspace*{-0.5cm} \noindent \rule{8.65cm}{.02cm}}
\def\resume{\hskip 9.3cm \rule{8.65cm}{.02cm} \begin{multicols}{2}\vspace*{-0.8cm} \noindent}
\def\umo{\rlap{\"{$\,$\ }} \,\,\,\kern -.3 cm o}

\newcount\hours \newcount\minutes \newcount\a \newcount\b
\def\gmt{\hours = \time \divide\hours by 60 \a =\hours \multiply \a by 60
\minutes = \time \advance \minutes by -\a
{\ifnum\hours<10 0\fi}\number\hours
{\ifnum\minutes<10 0\fi}\number\minutes}
\def\gday{\b=\year \advance \b by -1900
\number\b{\ifnum\month<10 0\fi}\number\month{\ifnum\day<10 0\fi}\number\day}
\def\today{\number\day\ 
\ifcase\month\or January\or February\or March\or April\or May\or 
June\or July\or August\or September\or October\or November\or December\fi
\ \number\year}

 \def\De{\Delta}

\def\casefr#1/#2 {\case{#1}{#2}}
\def\part{\partial} 

\def\half{\case{1}{2}}  

  \def\prop{\propto}

\def\sump#1{\lower .15in\hbox{$\stackrel{\displaystyle{\sum}'} {\scriptstyle #1}$}}

\def\tQ{\tilde Q} \def\ud{\De u} \def\tH{\tilde H} \def\fff{}
\def\resume{\hbox to 18cm {\hfill 
\rule{8.65cm}{.02cm}}\begin{multicols}{2}\vspace*{-0.8cm} \noindent}

\begin{document}

\title{\vbox to 0pt {\vskip -1cm \rlap{\hbox to \textwidth {\rm{\small FOR    
      PHYSICS OF FLUIDS} \hfill chao-dyn/9803037}}}Burgers Turbulence with 
Large-scale Forcing}

\author{Toshiyuki Gotoh$^{1,*}$ and Robert H. Kraichnan$^{2,\dag}$}

\address{$^1$Department of Systems Engineering, Nagoya Institute of 
Technology, Showa-ku, Nagoya 466, Japan}
\address{$^2$369 Montezuma 108, Santa Fe, NM 87501-2626}

\date{25 March 1998}

\abstract{Burgers turbulence supported by white-in-time random forcing at low 
wavenumbers is studied analytically and by computer simulation. The peak of 
the probability distribution function (pdf) $Q(\xi)$ of velocity gradient 
$\xi$ is at $\xi=O(\xi_f)$, where $\xi_f$ is a forcing parameter. It is 
concluded that $Q(\xi)$ displays four asymptotic regimes at Reynolds number 
$R\gg1$: (A) $Q(\xi) \sim \xi_f^{-2}\xi\exp(-\xi^3/3\xi_f^3)$ for $\xi \gg 
\xi_f$ (reduction of large positive $\xi$ by stretching); (B) $Q(\xi) \sim 
\xi_f^2|\xi|^{-3}$ for $\xi_f \ll -\xi \ll R^{1/2}\xi_f$ (transient inviscid 
steepening of negative $\xi$); (C) $Q(\xi) \sim |R\xi|^{-1}$ for $R^{1/2}\xi_f 
\ll -\xi \ll R\xi_f$ (shoulders of mature shocks); (D) very rapid decay of $Q$ 
for $-\xi \ge O(R\xi_f)$ (interior of mature shocks). The typical shock width 
is $O(1/Rk_f)$. If $R^{-1/2} \gg rk_f \gg R^{-1}$, the pdf of velocity 
difference across an interval $r$ is found be $P(\ud,r) \propto 
r^{-1}Q(\ud/r)$ throughout regimes A and B and into the middle of C.
}

\vskip .4cm

\begin{multicols}{2}

\Sec{INTRODUCTION}
A number of papers in the past several years have treated the physics of 
turbulent solutions of Burgers equation \cite{1,2,3,4,5,6,7,8,9,10,11,12,13}. 
A variety of results on statistics of solutions, sometimes contradictory, have 
been reported. Burgers turbulence driven by forcing that varies infinitely 
rapidly in time (white-in-time) was first systematically studied by Chekhlov 
and Yakhot \cite{2,6}. The present paper is concerned with  Burgers turbulence 
supported by white-in-time forcing that has a compact wavenumber spectrum.

Two processes act upon the velocity field injected by a forcing term: the 
self-advection of the velocity field steepens negative velocity gradients and 
reduces positive gradients. The viscous term relaxes the curvature of the 
velocity field. The effects of each term are easy to treat in isolation, but 
the combination of the two terms poses nontrivial difficulties.

At large Reynolds numbers, the self advection tends to produce sawtooth 
structures with smooth ramps of gentle positive velocity gradient and narrow 
shocks of strong negative gradient. The shock width is $O(\nu/\ud)$, where 
$\ud$ is the jump in velocity across the shock and $\nu$ is viscosity. Shocks 
can interact. A strong shock can move across the domain and swallow weaker 
structures in its path.

The present paper offers physically motivated approximations on terms in the 
exact equations of motion for the probability distributions of velocity 
gradient and velocity difference and examines their consistency. We assume 
that the forcing has compact spectral support at low wavenumbers and is white 
in time. The results are tested against computer simulations and the relation 
to other theoretical approaches is discussed. Four asymptotic regions are 
predicted at large Reynolds number: a region of large positive gradient where 
the gradient probability density decreases very rapidly; a region of 
intermediate negative gradient where the density follows a $-3$ law; a region 
of larger intermediate gradient where the density follows a $-1$ law; and 
finally an outer region of negative gradient where decay is very rapid. The 
power-law regions are mediated by transient advective steepening of gradients 
and the shoulders of mature shocks, respectively, while the outer region 
represents gradients within strong shocks.

\Sec{STATISTICAL EQUATIONS FOR THE VELOCITY GRADIENT}

Burgers equation with forcing is
\opn
u_t + uu_x = \nu u_{xx} + f,
\qno;1;
where $u(x,t)$ is a one-dimensional velocity field and $f(x,t)$ is a forcing 
term. The left side of \rf;1; is the Lagrangian time derivative of $u$, 
measured along a fluid-element trajectory. We shall assume that the forcing is 
white in time, statistically homogeneous and stationary, with compact spectral 
support concentrated about a wavenumber $k_f$. Let
\opn
B = \int_0^t\av{f_x(x,t)f_x(x,s)}ds,
\qno;2;
where $u(x,t=0)=0$ and $\av{\ }$ denotes ensemble average. A characteristic 
forcing strain rate and Reynolds number induced by the forcing may be defined 
by $\xi_f=B^{1/3}$ and $R=\xi_f/(\nu k_f^2)$. The steady-state values of rms 
velocity $u_{rms} = \av{u^2}^{1/2}$ and typical shock jump induced by the 
forcing are both $O(\xi_f/k_f)$. The typical shock width is $\nu/u_{rms}$ and 
the typical velocity gradient $\xi=u_x$ within a shock is $R\xi_f$.

Differentiation of \rf;1; yields
\opn
\xi_t + u\xi_x = -\xi^2 + \nu\xi_{xx} + f_x.
\qno;3;
The $\xi^2$ term in \rf;3; represents advective gradient intensification or 
diminution along Lagrangian trajectories. An exact equation of motion for the 
probability distribution function (pdf) $Q(\xi)$ of $\xi$ follows from \rf;3;:
\opn
{\part Q\over\part t} - {\part\over\part\xi} \left( \xi^2Q \right) = 
-\nu{\part[H(\xi)Q(\xi)]\over\part\xi} + \xi Q +B{\part^2Q\over \part\xi^2}.
\qno;4;
Here $H(\xi) \equiv \av{\xi_{xx}|\xi}$ denotes the ensemble mean of 
$\xi_{xx}$ conditional on fixed $\xi$. This relation was derived in \cite{1} 
for zero forcing by following probabilities along Lagrangian trajectories. The 
$B$ term in \rf;4; expresses in standard fashion the outward diffusion of 
probability due to white-in-time forcing.

The $\xi^2$ term in \rf;3; plays two opposed roles. If $\xi < 0$, it 
intensifies the gradient but, at the same time, squeezes the fluid and thereby 
decreases the measure along $x$ associated with an interval $d\xi$. If $\xi > 
0$, the gradient is decreased but measure is increased by stretching of the 
fluid. The intensification or diminution of gradient is expressed in \rf;4; by 
the $\part(\xi^2Q)/\part\xi$ term and the rate of change of measure is 
expressed by the $\xi Q$ term. An identity for homogeneous fields,
\opn
{\part\over\part\xi} \left( \avv{u\xi_x|\xi} Q\right) \equiv \xi Q,
\qno;5;
gives an alternative expression for the rate of change of measure.

\Sec{APPROXIMATIONS FOR ASYMPTOTIC RANGES}

The limit $R \to \infty$ suggests the existence of several asymptotic ranges 
of $Q(\xi)$, whose form can be found if physically-based approximations to the 
terms of \rf;4; are valid. We shall outline these approximations compactly in 
this Section, and later address some questions and paradoxes that arise.

First, consider places in the flow, at a given time, where forcing has 
produced an extraordinarily large positive value $\xi \gg \xi_f$. Stretching 
should very quickly flatten such regions, suggesting that the $\nu$ terms in 
\rf;3; and \rf;4; can be neglected in comparison to the other terms in a 
statistically steady state. The $Q$ equation then reduces to 
\opn
B{\part^2Q\over\part\xi^2} + \xi^2{\part Q\over\part\xi} + 3\xi Q = 0,
\qno;6;
The general solution involves $\exp(-\xi^3/3B)$. The solution that vanishes 
at $\xi = +\infty$ is 
\opn
Q(\xi) \approx C_+\,\xi_f^{-2}\xi\exp(-\xi^3/3B) \quad (\xi \gg \xi_f),
\qno;7;
where $C_+$ is a dimensionless constant and the inequality defines the range 
in which we hope the neglect of viscous effects is justified. The exponential 
factor in \rf;7; was first found by Polyakov \cite{4}, but with a different 
prefactor.

Next, consider places where $\xi < 0$ and $\xi_f \ll -\xi \ll R\xi_f$. The 
first inequality implies that the gradient is steepening with a time constant 
$\ll 1/\xi_f$ and the second inequality makes $|\xi|$ small compared to 
typical gradients found within a shock. We therefore neglect the $\nu$ and $B$ 
terms in \rf;4;, yielding in steady-state
\opn
\xi^2{\part Q\over\part\xi} + 3\xi Q = 0.
\qno;8;
The solution of physical interest is
\opn
Q(\xi) \approx C_{-3}\,\xi_f^2|\xi|^{-3} \quad (\xi_f\ll -\xi \ll 
R^{1/2}\xi_f),
\qno;9;
where $C_{-3}$ is another dimensionless constant. Again, the inequalities 
express the range in which we hope our approximations are valid. The second 
inequality stated in \rf;9; is stronger than called for by the argument just 
given. This is because the transient steepening contribution \rf;9; to $Q$ is 
overshadowed in the range $R^{1/2}\xi_f \ll -\xi \ll R\xi_f$ by contributions 
from the shoulders of the strong, quasi-equilibrium shocks that decay with 
time constants $O(1/\xi_f)$.

There is a simple physical explanation for \rf;9;. Consider a place where 
forcing has produced negative $\xi$ with $-\xi=O(\xi_f)$ over a region of size 
$O(1/k_f)$ where $\xi_{xx}$ is small enough that viscous effects are 
negligible. Steepening of the gradient will carry the gradient to larger 
negative values and the time in which it is $O(|\xi|)$ will be $O(1/|\xi|)$. 
Simultaneously, the measure will shrink to $O(\xi_f/|\xi|)$. Thus this process 
will contribute $O(1/|\xi|^2)$ to $|\xi|Q(\xi)$, the probability of finding 
the gradient at $O(|\xi|)$. This implies $Q(\xi) \prop |\xi|^{-3}$.

An equilibrium single-shock solution of \rf;1;, \rf;3; is $u(x)=-u_s\tanh(u_s 
x/2\nu)$,
\opn
\xi(x)=-\left({u_s^2\over2\nu}\right) {\rm sech}^2\left({u_s x\over2\nu} 
\right).
\qno;10;
The jump across the shock is $2u_s$. The measure on $x$ of places with 
gradient between $\xi$ and $\xi+d\xi$ is $d\xi/|\xi_x|$. By differentiation of 
\rf;10;, the contribution of this shock to $Q(\xi)$ is then
\opn
Q(\xi) \prop {\nu\over|\xi|\sqrt{u_s^2-2\nu|\xi|}}$ \quad ($|\xi| < 
u_s^2/2\nu).
\qno;11;
If $u_s = O(u_{rms})$ and $|\xi| \ll R\xi_f$, this is simply $Q(\xi) \prop 
\nu/u_s|\xi|$.

Now consider a train of such shocks, spaced $1/k_f$ apart, with a uniform 
gradient $2u_sk_f$ added so that there is no secular change of $u$. It follows 
that the normalized total contribution to $Q(\xi)$ for $\xi_f \ll |\xi| \ll 
R\xi_f$ is $O(1/R|\xi|)$. The significance of $R$ here is that it measures the 
ratio of shock spacing to shock width. For $|\xi| \ll R^{1/2}\xi_f$, this 
contribution to $Q$ is overpowered by \rf;9;, but for $|\xi| \gg 
R^{1/2}\xi_f$, it overpowers \rf;9;. Thus the third asymptotic range is
\opn
Q(\xi) \approx {C_{-1}\over R|\xi|} \quad (R^{1/2}\xi_f \ll -\xi \ll R\xi_f),
\qno;12;
where $C_{-1}$ is yet another dimensionless constant. This result is 
unchanged in form if the shocks of fixed jump $2u_s$ are replaced by shocks 
with a distribution of jump values $u_s = O(u_{rms})$.

The form \rf;12; may also be obtained by approximate dynamic analysis that 
follows the evolution of those regions of negative gradient that evolve into 
the shoulders of an equilibrium shock. An example is the mapping closure for 
Burgers shocks \cite{1}.

Equation \rf;11; may be integrated over a distribution $P_s(u_s)$ of shock 
jumps to get the form of $Q(\xi)$ for $-\xi \ge O(R\xi_f)$, where 
contributions from within the shocks must dominate $Q$. Thus,
\opn
Q(\xi) \approx {C_su_{rms}\over R|\xi|} \int_{\sqrt{2\nu|\xi|}}^\infty 
{P_s(u_s)du_s\over\sqrt{u_s^2-2\nu|\xi|}} \quad (-\xi \ge O(R\xi_f)),
\qno;13;
where $C_s$ is a dimensionless constant. $P_s(u_s) \prop u_s\exp(-\half 
au_s^2/u_{rms}^2)$, where $a$ is a dimensionless constant, is one of the forms 
for which the integration in \rf;13; is analytic. It yields
\opn
Q(\xi) \approx {C_s\over R|\xi|}\exp(-a|\xi|R/\xi_f) \quad (-\xi \ge 
O(R\xi_f)).
\qno;14;
It happens that the form \rf;14; is also obtained by applying mapping closure 
to \rf;1; and \rf;3;, both in the case of free decay \cite{1} and in the 
present forced case.

Note that \rf;14; gives an exponential fall-off while the assumed $P_s(u_s)$ 
falls off as a Gaussian. This is because the gradient in the center of a shock 
is $-u_s^2/2\nu$. The shock width $\nu/u_s$ decreases as $u_s$ increases.

Several authors have concluded that the tail of the pdf of shock jump falls 
off more rapidly than that of the Gaussian forcing; in particular, $\ln 
P_s(u_s) \prop -(u_s/u_{rms})^3$ at large $u_s$ \cite{3,8,9}. The reason is 
that the decay rate of a shock is $O(u_sk_f)$, which increases with increase 
of $u_s$. The corresponding prediction for the gradient pdf is $\ln Q(\xi) 
\prop -(|\xi|R/\xi_f)^{3/2}$ at large negative $\xi$, instead of \rf;14;.

\Sec{THE CONDITIONAL MEAN OF DISSIPATION}

The conditional mean $H(\xi)$ embodies all the difficulty in solving \rf;4;. 
We wish now to examine whether this mean behaves in a way that realizes the 
asymptotic $R \to \infty$ ranges of $Q(\xi)$ proposed in Sec. III.

The steady-state asymptotic behavior of $H(\xi)$ at $-\xi \gg R\xi_f$ can be 
inferred immediately if the $B$ term can be neglected in that range. $H$ must 
balance the remaining two terms, which represent steepening and loss of 
measure:
\opn
\nu H(\xi) \approx \xi^2 + {1\over Q(\xi)} \int_{-\infty}^\xi 
Q(\xi')\xi'd\xi'.
\qno;15;
Note that statistical homogeneity requires $\int_{-\infty}^\infty Q(\xi)\xi 
d\xi=0$. If $Q(\xi)$ falls off faster than algebraically as $\xi \to -\infty$, 
then the $\xi^2$ term in \rf;15; is dominant. The integration in \rf;15; is 
analytic for the example \rf;14;, giving
\opn
\nu H(\xi) \approx \xi^2 + R\xi_f\xi/a.
\qno;16;

Paradoxically, $\nu H(\xi)$ plays an important role in \rf;4; at $|\xi| = 
O(\xi_f)$, however small $\nu$ may be. This is because of the contribution 
from the shoulders of mature shocks. If this is so, how can the $H$ term be 
negligible for larger negative $\xi$, as assumed in obtaining \rf;9;?

Differentiation of \rf;10; gives
\opn
\nu\xi_{xx} = 3\xi^2 + (u_s^2/\nu)\xi.
\qno;17;
If $u_s=O(u_{rms})$, the second term in \rf;17; is $O(R\xi_f\xi)$, and it 
dominates the first term for $|\xi| \ll R\xi_f$.  Consider the measure, per 
unit length of $x$, of the portion of a shock shoulder where $|\xi|=O(\xi_f)$. 
If the shock spacing is $O(1/k_f)$, this measure is $M_S(|\xi|=O(\xi_f)) = 
O(R^{-1})$ while the total measure per unit length $M_T(|\xi|=O(\xi_f))$, 
where $|\xi|=O(\xi_f)$, is $O(1)$. The total measure is dominated by field, 
freshly injected by the forcing, that has not interacted with shocks. If $R$ 
is large, $\nu H(\xi=O(\xi_f))Q(\xi=O(\xi_f))$ is dominated by the 
contribution $(u_s^2/\nu)\xi=O(R\xi_f^2)$ from the shock shoulders. The 
corresponding contribution to $\nu H(\xi)$ is $O(R\xi_f^2)M_S/M_T$. Since 
$M_S/M_T=O(R^{-1})$, this implies that the shock shoulders make an 
$O(R^{-1}R\xi_f^2)$ contribution to $H(\xi)$ for $\xi$ negative and 
$O(\xi_f)$. Thus the $\nu$ term in \rf;4; is the same order as the 
$\part(\xi^2Q)/\part\xi$ term for such $\xi$. It is independent of $R$ as 
$R\to\infty$.

Next consider the putative range \rf;9;. Here the dominant part of $\xi_{xx}$ 
within the shock shoulders is again $\prop \xi$, while by \rf;12; the 
contribution of the shoulders to $Q(\xi)$ is $\prop 1/\xi$. The contribution 
of the shoulders to $H(\xi)Q(\xi)$ is again the dominant one. This 
contribution thus goes like $\xi\xi^{-1}$ and is independent of $\xi$ to 
leading order. The $\nu$ term in \rf;4; vanishes to leading order when the 
$\xi$ differentiation is performed. This is consistent with the neglect of the 
$\nu$ term in deriving \rf;9;.

The regions outside the shoulders have negligible $\nu\xi_{xx}$ at large $R$. 
A relative-measure estimate, like that above, then implies $H(\xi) \prop 
\xi^3$ for $\xi_f \ll -\xi \ll R^{1/2}\xi_f$. $H(\xi)$ grows with $|\xi|$ 
throughout the range \rf;9;, but neglect of $\part[H(\xi)Q(\xi)]/\part\xi$ 
appears to be valid.

\Sec{STATISTICAL EQUATIONS FOR VELOCITY DIFFERENCES}

Let $P(\ud,r)$ be the pdf of $u(x+r,t)-u(x,t)$, where $r$ is always taken 
positive. If $z=\ud/r$, then $P(\ud,r) = \tQ(z,r)/r$ where $\tQ$ is the pdf of 
$z$. An equation of motion for $P$ is obtained by following Lagrangian 
trajectories as in the derivation \cite{1} of \rf;4;. Two trajectories must be 
followed simultaneously in the present case. The result is
\[{\part P\over\part t} = -\nu {\part\over\part\ud}\left[ \av{\left( 
{\part^2\over\part x^2} + {\part^2\over\part x'^2} \right)\ud \biggr|  \ud} P 
\right] - \ud {\part P\over\part r} \nonumber \]
\opn
+ \av{\left( {\part\ud\over\part x'} - {\part\ud\over\part x} \right) \biggl| 
\ud}P + F(r){\part^2 P\over\part\ud^2},
\qno;18;
where $F(r)=\int_0^t[f(x+r,t)-f(x,t)][f(x+r,s)-f(x,s)]ds$ and $x'=x+r$. The 
next-to-last term in \rf;18; expresses the change of measure at the points $x$ 
and $x'$. It contains $\part u(x,t)/\part x$ and $\part u(x',t)/\part x'$. 
Instead of the $\xi^2Q$ gradient intensification term in \rf;4;, there is now 
the $\part P/\part r$ term. It expresses the change of the label $r$ carried 
by the velocity difference between the two fluid elements as they move closer 
($\ud < 0$) or apart ($\ud > 0$).

An equation for $\tQ$ equivalent to \rf;18; may be written as
\[
{\part \tQ\over\part t} = r^{-2}F(r){\part^2\tQ\over\part z^2} + 
z^2{\part\tQ\over\part z} + z\tQ -zr{\part \tQ\over\part r} \nonumber \]
\opn
+ M(z,r)\tQ - \nu{\part[\tH(z,r)\tQ]\over\part z},
\qno;19;
where $z=\De u/r$,
\[
M(z) = 2z + 2r\av{\part z/\part r|z}, \nonumber \]
\opn
\tH(z) = \av{u_{xx}(x',t)-u_{xx}(x,t)|z}/r.
\qno;20;
In \rf;20;, $M$ has been re-expressed by use of the identity $\part{\bar 
u}/\part{\bar x} \equiv \part\ud/\part r$, where $\bar u=\half[u(x,t) + 
u(x',t)]$. $M$ can be further tranformed \cite{14} by use of the 
homogeneous-field identity \cite{15}
\opn
\av{\part z/\part r|z} \equiv - {1\over\tQ(z,r)} \int_{-\infty}^z 
{\part\tQ(z',r)\over r}dz'.
\qno;21;
This expresses the advective contributions to \rf;19; entirely in terms of 
$\tQ$, at the expense of creating an integro-differential equation. An 
equation equivalent to \rf;19; was obtained by Polyakov \cite{4} by exploiting 
translation invariance of the equation of motion for the characteristic 
function (Fourier transform) of $\tQ$.

If the limit $r\to0$ is taken, $z \to \xi$ and $\tQ(z,r) \to Q(\xi)$, which 
is independent of $r$. Also, $\tH(z,r) \to H(\xi)$ and $r^{-2}F(r) \to B$. 
Then \rf;19; goes into \rf;4;.

If $r$ is greater than typical shock widths and $z$ is large enough, the 
right side of \rf;21; is dominated by contributions from shocks such that one 
of the two points $x$ and $x'$ lies in the shock front while the other lies 
outside. The integration limits in \rf;21; express the fact that all shocks 
with jumps $\ge \ud$ contribute.

The pdf of $z=\ud/r$ should approximate that of $\xi$ if, in the 
line-segments of length $r$ that contribute significantly at a given $z$, 
$\xi$ fluctuates negligibly over the distance $r$. For $\xi > 0$ or $|\xi| 
\sim \xi_f$, a plausible sufficient condition for this is $rk_f \ll 1$. For 
$-\xi \gg \xi_f$, the typical spatial scale of variation of $\xi$ has been 
decreased by squeezing to $\sim \xi_f/k_f|\xi|$. The requirement for 
negligible variation of $\xi$ over $r$ is then $r \ll \xi_f/k_f|\xi|$, which 
can be rewritten as $rk_f \ll \xi_f/|\xi|$ or $r|\xi| \ll u_{rms}$. This 
excludes $\ud$ values due to the presence of a typical shock within $r$. The 
condition for the pdfs of $z$ and $\xi$ to fall on each other throughout the 
$-3$ range is consequently $rk_f \ll R^{-1/2}$. If $R^{-1/2} \gg rk_f \gg 
R^{-1}$, the collapse of the pdfs of $z$ and $\xi$ should extend into the $-1$ 
range. In general, if $rk_f \ll 1$ the pdf of $z$ should fall below that of 
$\xi$ at negative $\xi$ large enough that $rk_f \ll \xi_f/|\xi|$ is violated.

\Sec{COMPARISON WITH SIMULATIONS}

Computer solution of \rf;1; by an SX4 machine was carried out on a cyclic 
domain of $N=2^{17}$ to $N=2^{20}$ points with unit spacing. The forcing had a 
wavenumber spectrum of form (A) $k^2\exp[-(k/k_f)^2]$ or (B) 
$k^4\exp[-(k/k_f)^2]$. The forcing field at each time step was an independent 
realization of Gaussian statistics. The initial spectrum for $u$ had the form 
$k^2\exp[-(k/k_f)^2]$, with variance chosen to minimize transients in 
$u_{rms}$. The simulations were continued until the statistics of interest 
were stationary ($\sim 10^5$ time steps). The steady state Reynolds numbers 
$R=u_{rms}/\nu k_f$ ranged from 15 to 18000. The integrations were performed 
using second-order schemes in $x$ and $t$. Table I shows the simulation 
parameters for the runs reported here. $\av{R}$ is the approximate time 
average of $R$ over the period in which $Q(\xi)$ exhibited a statistically 
stationary state. Statistics were averaged over sets of similar runs that 
differed only in the random numbers used in realizing the forcing and initial 
velocity spectra. For Runs 3--5, additional averaging was performed over time 
in the statistically stationary state.

Fig.~\fff1 shows $\log_{10}[\av{\xi^2}^{1/2} Q(\xi)]$ plotted against 
$\xi/\av{\xi^2}^{1/2}$ for $R \sim 15, 1200, 18000$ (Runs 3--5). Note the 
increase in sharpness of the peak as $R$ increases. Fig.~\fff2 shows the 
central region of $\xi_fQ(\xi)$ plotted against $\xi/\xi_f$ for three runs 
with forcing spectrum (B), $R \sim 15, 1200, 18000$ (Runs 3--5) and Run 1, 
with forcing spectrum (A) at $R \sim 15$. With the $\xi_f$ scaling, the 
central region is substantially insensitive to change of forcing spectrum or 
$R$.

Fig.~\fff3 shows superimposed plots, for $R \sim 1200$ (Run 4) and $R \sim 
18000$ (Run 5), of $\log_{10}[\xi_f Q(\xi)]$ against $\log(|\xi|/\xi_f)$ for 
$\xi < 0$. The straight lines have slopes of $-3$ and $-1$. The $R \sim 18000$ 
data seem consistent with the existence of the proposed asymptotic ranges 
\rf;9; and \rf;12; but clearly even higher $R$ values would be needed to make 
an unambiguous case.

The steady-state pdf equation, obtained by setting $\part Q/\part t=0$ in 
\rf;4; has interesting stability properties. If the dissipation term $\nu\part 
[H(\xi)Q(\xi)]/\part\xi$ is omitted, the general solution is analytically 
accessible. It contains exponential factors that make numerical solution 
violently unstable if carried out in the negative $\xi$ direction and highly 
stable if carried out in the positive $\xi$ direction. A consequence is that 
the central peak of $Q$ is quite insensitive to the form assigned to the $\nu$ 
term.

Fig.~\fff4 shows the central part of $\xi_fQ(\xi)$ plotted against 
$\xi/\xi_f$ for four cases: (a) the $R \sim 1200$ simulation (Run 4); (b) the 
left-to-right numerical solution of \rf;4; in steady state with dissipation 
term set to zero; (c) the left-to-right numerical solution with dissipation 
term taken as $0.45\xi_fQ(\xi)$; (d) the left-to-right solution with 
dissipation term taken as $0.8609\xi_fQ(\xi)/(1+\xi^2/\xi_f^2)$. The three 
numerical solutions were started at large negative $\xi$, where $Q(\xi) \prop 
\xi^{-3}$ and normalized to unit probability.

The first notable thing about Fig.~\fff4 is that case (b) is so close to case 
(a). It is has a small, unphysical negative region at $\xi > 0$, but 
nevertheless lies near (a) over the peak region. The forms (c) and (d) for the 
dissipation are even closer to (a). They decay according to \rf;7; as 
$\xi\to\infty$.

Case (c) is the case $a=0$ of the closure approximation 
$\nu\part[HQ]/\part\xi = (a\xi+b\xi_f)Q$ introduced by Polyakov \cite{4} and 
studied also by Boldyrev \cite{12}. This closure was obtained as a low-order 
truncation of the operator-product expansion associated with Burgers equation.

For our present purposes, (c) and (d) are simply two functional forms that 
serve to increase $\part^2Q/\part\xi^2$ for $|\xi| \sim \xi_f$. The choice of 
a functional form for this purpose is only weakly constrained, but the 
numerical coefficient is not. It is fixed by the requirements that $Q(\xi)$ be 
positive for all $\xi$ and decay faster than algebraically as $\xi \to \infty$ 
\cite{4,12}. The second requirement is satisfied also by case (b). If the 
numerical coefficients in cases (c) and (d) are increased, the right tail of 
$Q(\xi)$ decays like $\xi^{-3}$ instead of like \rf;7;.  If the coefficients 
are decreased, a negative region like that in case (b) is produced. The 
numbers 0.45 and 0.8609 stated above are approximations to the exact marginal 
values.

It was noted in Sec. V that viscous effects from the shoulders of mature 
shocks are present at $|\xi| \sim \xi_f$ even as $R \to \infty$. Fig.~\fff4 
suggests that the principal consequence of these effects in the $Q$ equation 
are not upon the form of the central peak, whose shape is remarkably stable, 
but upon the decaying region of $Q$ at larger positive $\xi$.

The putative asymptotic range \rf;7; is difficult to define well by 
simulation. $Q(\xi)$ decays so rapidly with increase of $\xi$ that very large 
sample sets are needed. Fig.~\fff5 is a plot of $3B\part(\ln Q)/\part(\xi^3)$ 
vs $\xi^3/B$ for Run 2, a $R \sim 15$ simulation with forcing spectrum (B). 
The $\exp(-\xi^3/3B)$ factor in \rf;7; seems supported. But the approach to 
the horizontal asymptote is protracted, consistent with the presence of a 
positive-power prefactor.

A definitive test of the prefactor exponent in \rf;7; is more difficult. 
Fig.~\fff6 is a plot of $3\part(\ln Q)/\part(\ln\xi^3)$ against $\xi^3/B$. If 
the $\exp(-\xi^3/3B)$ factor is present in $Q$, the prefactor exponent is 
given by the intercept at $\xi^3=0$ of a straight line drawn through this plot 
at large $\xi^3$.

In order to help resolve the prefactor, we have included in Figs.~\fff5 and 
\fff6 curves corresponding to a mapping approximation \cite{1} carried out 
with the same $\nu$ and forcing parameters as the simulation. A detailed 
description of the approximation for the forced case will be given elsewhere. 
We report now is that it yields \rf;7; at infinite $R$ but gives $Q(\xi) \prop 
\xi_f^{-2}\xi\exp(-{\rm const}\,\nu\xi/u_{rms}^2 - \xi^3/3B)$ for $\xi \gg 
\xi_f$ at finite $R$. The simulation and mapping curves in Fig. \fff6 lie 
close to each other. Straight lines drawn through the outer parts of both 
curves ($10 \le \xi^3/B \le 15$) intercept the vertical axis near 1.

It is also difficult in the simulations to resolve the conditional mean 
$H(\xi)$ cleanly for large negative $\xi$. In addition to the need for large 
sample sets, the $x$ grid must be sufficiently fine to resolve unusually 
narrow shocks. Fig.~\fff7 shows $\nu H(\xi)/\xi_f^2$ plotted against 
$\xi/\xi_f$ ($\xi<0$) for Run 3, which has high resolution. Also shown are the 
parabolas $\nu H(\xi)=\xi^2$ and $\nu H(\xi) = \xi^2+\xi_C\xi$, where $\xi_C$ 
is the value at which the simulation data for $H(\xi)$ change sign. The latter 
function is an approximation suggested by \rf;17;. The second term in the 
asymptotic relation \rf;15; obviously is negative for $\xi < 0$. This implies 
that $\xi^2$ is an upper bound to $\nu H(\xi)$ for large negative $\xi$.

Fig.~\fff8 shows data for $\xi_frP(\ud,r)$ from the $R \sim 1200$ simulation 
(Run 4) plotted against $\ud/r\xi_f$ for a number of values of $r$. Also shown 
is $\xi_fQ(\xi)$ plotted against $\xi/\xi_f$. As $r$ decreases, the curves for 
$P$ follow that for $Q$ over an increasingly long range of $\ud/r$. At the 
smaller $r$ values, this collapse extends into the $-1$ range of the $Q$ 
curve. The envelope of the knees of the $P$ curves, beyond the region of 
collapse, follows a line of slope $-2$ on the log-log plot. This is because 
the regions of large $\ud/r$ are dominated by contributions from the mature 
shocks, and scale with $u_{rms}$. The latter scaling is clearly shown in 
Fig.~\fff9, where $P(\ud,r)/\xi_fr$ is plotted against $\ud/u_{rms}$. The 
outer regions of the curves collapse.

Figs.~\fff10 and \fff11 are similar to Figs.~\fff8 and \fff9 except that they 
show data for $R \sim 18000$ (Run 5).



\Sec{Discussion}

The theory of Burgers equation with low-wavenumber, white-in-time forcing has 
had a variety of treatments, and there has been a variety of predictions. The 
$\exp(-\xi^3/3B)$ factor in the right tail of $Q(\xi)$ seems generally 
accepted. It was first predicted by Polyakov \cite{4} on the basis of the 
truncation $\nu\part[HQ]/\part\xi = (a\xi+b\xi_f)Q$ of the operator product 
expansion, as noted in Sec. VI. Later it was recovered by instanton analysis 
\cite{8,9,10}.

Equation \rf;4; is exact. If the $\nu$ term in \rf;4; can be neglected in the 
right tail, it follows that the prefactor is $\xi$, as shown in \rf;7;. 
Neglect of viscous effects in the right tail is plausible but not obviously 
justified. One must consider the effects of possible proximity of regions of 
large positive $\xi$ to the shoulders of strong shocks.

All values $a \ne 0$ in Polyakov's closure give significant viscous effects 
in the right tail, with the consequence that the predicted prefactor is 
$\xi^{1-a}$. Instanton analysis that neglects viscous effects at the start 
\cite{10} must yield the $\xi$ prefactor. However, extraction of the prefactor 
from instanton analysis is a delicate matter that requires careful treatment 
of fluctuations about the saddle point.

There is also disagreement about the exponent of the putative power-law range 
for intermediate negative $\xi$. The predictions have ranged from $-2$ 
\cite{6,7} to $-7/2$ \cite{11}. Polyakov's closure gives $q=3-a$; values $0 
\le a \le 1$ have been considered \cite{4,12}.

As discussed in Sec. IV, the dissipation contribution from the shoulders of a 
distribution of ideal Burgers shocks gives $H(\xi) \prop |\xi|^3$ if $Q(\xi) 
\prop |\xi|^{-3}$, so that the $\nu$ term in \rf;4; vanishes to leading order, 
as required for consistency. If $Q(\xi) \prop |\xi|^{-q}\ \ (q \ne 3)$, \rf;4; 
in steady state requires
\opn
\nu H(\xi) \approx {3-q \over 2-q}\xi^2.
\qno;22;
Thus $H(\xi)$ is negative if $2 < q < 3$ and positive if $q > 3$ or $q < 2$. 
If $R$ is large, contributions to $H$ for the range $|\xi| \ll R\xi_f$ in 
question can come only from the shoulders of shocks. The profile of $\xi$ in a 
shock is a negative peak so that, whatever the precise profile, the curvature 
of the shoulders is negative and the contribution to $H(\xi)$ therefore is 
negative. This appears to rule out $q > 3$ and $q < 2$. Values $q \ne3$ imply 
shock profiles different from that of the ideal Burgers shock \rf;10;.

Values $q \le 2$ are ruled out for another reason. Homogeneity requires 
$\int_{-\infty}^\infty Q(\xi)\xi\,d\xi=0$. The contribution to this integral 
from $\xi > 0$ is finite and independent of $R$ as $R \to \infty$, if the 
right tail of $Q$ falls off rapidly in the fashion generally accepted. But if 
the negative-$\xi$ power-law range extends from $|\xi| = O(\xi_f)$ to $|\xi| = 
O(R^c\xi_f)$ where $c > 0$, then the contribution of this range to the 
integral is negative and becomes infinite as $R\to\infty$. Since the 
contribution from the left tail beyond the powerlaw range is also negative, 
the homogeneity condition cannot be satisfied in the limit.

The value $q=7/2$ was proposed in the course of an analysis of the inviscid 
Burgers equation \cite{11}. One thing established rigorously in this work is 
that the mean spacing of strong shocks is $O(1/k_f)$ if the forcing has 
compact spectral support about $k_f$. The value $q = 7/2$ was obtained by 
examining the behavior of $u(x,t)$ in the immediate vicinity of the formation 
of an incipient shock, and then looking back in time at the structure of the 
regions that were destined to form these vicinities. We believe that this 
procedure yields a biased sample of the pre-shock velocity field. Equation 
\rf;9; is intended to describe the probability balance associated with the 
evolution of all regions of negative $\xi$. Only a zero-measure set of these 
regions is destined to form part of the immediate vicinity of an incipient 
shock in the limit $R\to\infty$.

One effect not taken into account in the derivation of \rf;9; is passage of a 
mature shock through a region where $\xi$ is steepening. Such passage can wipe 
out the local process. Since negative $\xi$ increases in magnitude until it 
either forms part of a shock or is wiped out by a shock collision, the 
collision process should make $Q(\xi)$ fall off more rapidly than \rf;9; with 
increase of $|\xi|$. The rate of increase of $\xi$ within a fluid element, 
before viscous effects are felt and neglecting forcing effects, is $\xi(t) 
\prop \xi_0/(1-|\xi_0|t)$, where $t$ is measured from some effective initial 
time when $\xi=\xi_0$. The times at which $|\xi|$ is large compared to 
$|\xi_0|$ are all crowded into an interval $\ll 1/|\xi_0|$ at $t = 
O(1/|\xi_0|)$. The mean time between shock collisions is $O(1/k_fu_{rms})$. 
Because of the crowding in time, the the collision probability changes very 
slowly with $|\xi|$ for $|\xi| \gg \xi_f$, and we do not expect the collision 
effects to change the power law in \rf;9;.

\vskip 4mm

We are indebted to S. A. Boldyrev, S. Chen, A. Chekhlov, M. Chertkov, G. 
Falkovich, V. Lebedev, A. M. Polyakov, B. Shraiman, Y. Sinai, and V. Yakhot 
for valuable discussions and correspondence. T.G.'s work was supported in part 
by Grant-in-Aid for Scientific Research C,09640260 from the Ministry of 
Education, Science, Sports and Culture of Japan, and by the Computational 
Science and Engineering, Research for the Future Program, Japan Society for 
the Promotion of Science. R.H.K.'s work was supported by the United States 
Department of Energy, Office of Basic Energy Sciences, under Grant 
DE-FG03-90ER14118.

\end{multicols}

\vbox{
\hskip -0.5cm 
\begin{table}[h]
\hskip 2.0cm
\caption{SIMULATION PARAMETERS}
\vspace{2mm}
\begin{tabular}{cccccccccc} \\ 
Run & $\av{R}$ & $u_{rms}(t=0)$ & Spectrum & $k_f$ & $\nu$ & $N$ & $dt$ &
    $B$  & Set size \\ \hline
1& $15$ & $1$ & A & $0.02$& $10/3$ & $2^{17}$& $0.1$& 
   $4\times 10^{-6}$& $102$ \\
2& $15$ & $1$ & B & $0.02$& $10/3$ & $2^{17}$& $0.1$& 
   $7.2\times 10^{-6}$& $102$ \\
3& $15$ & $1$ & B & $0.02$& $10/3$ & $2^{20}$& $0.1$& 
   $7.2\times 10^{-6}$& $102$ \\
4& $1200$ & $1$ & B & $5\times 10^{-4}$ & $2$ & $2^{18}$ & $0.1$ & 
   $2\times 10^{-10}$ & $102$ \\
5& $18000$ & $1$  & B & $5\times 10^{-5}$ & $2.2$ & $2^{20}$ & $0.2$&
   $5\times 10^{-13}$ & $61$ \\ 
\end{tabular}
\end{table}
}

\begin{multicols}{2}

\opbib

\vbox to -.6in { }

\bibitem[*]{0} Electronic address: gotoh@system.nitech.ac.jp

\bibitem[\dag]{0} Electronic address: rhk@lanl.gov

\bb{1} T. Gotoh and R. H. Kraichnan, ``Statistics of decaying Burgers 
Turbulence,'' Phys. Fluids A \f.5., 445 (1993); T. Gotoh, ``Inertial range 
statistics of Burgers turbulence,'' Phys. Fluids \f.6., 3985 (1994).

\bb{2} A. Chekhlov and V. Yakhot, ``Kolmogorov turbulence in a 
random-force-driven Burgers equation,'' Phys. Rev. E, \f.51., 2739 (1995); A. 
Chekhlov and V. Yakhot, ``Kolmogorov turbulence in a random-force-driven 
Burgers equation: anomalous scaling and probability density functions,'' Phys. 
Rev. E \f.52., 5681 (1995).

\bb{3} M. Avellaneda, R. Ryan, and Weinan E, ``PDFs for velocity and velocity 
gradients in Burgers turbulence,'' Phys. Fluids \f.7., 3067 (1995).

\bb{4} A. M. Polyakov, ``Turbulence without pressure,'' Phys. Rev. E \f.52., 
6183 (1995).

\bb{5} J.-P. Bouchaud, M. M\'ezard, and G. Parisi, ``Scaling and 
intermittency in Burgers turbulence,'' Phys. Rev. E \f.52., 3656 (1995).

\bb{6} V. Yakhot and A. Chekhlov, ``Algebraic tails of probability functions 
in the random-force-driven Burgers turbulence,'' Phys. Rev. Lett. \f.77., 3118 
(1996).

\bb{7} J.-P. Bouchaud and M. M\'ezard, ``Velocity fluctuations in forced 
Burgers turbulence,'' Phys. Rev. E \f.54., 5116 (1996).

\bb{8} G. Falkovich, I. Kolokolov, V. Lebedev, and A. Migdal, ``Instantons 
and intermittency,'' Phys. Rev. E \f.54., 4896 (1996).

\bb{9} E. Balkovsky, G. Falkovich, I. Kolokolov, and V. Lebedev, 
``Intermittency of Burgers turbulence,'' Phys. Rev. Lett. \f.78., 1452 (1997).

\bb{10} V. Gurarie and A. Migdal, ``Instantons in the Burgers equation,'' 
Phys. Rev. E \f.54., 4908 (1996).

\bb{11} Weinan E, K. Khanin, A, Mazel, and Y. Sinai, ``Probability 
distribution functions for the random forced Burgers equation,'' Phys. Rev. 
Lett. \f.78., 1904 (1997).

\bb{12} S. A. Boldyrev, ``Velocity-difference probability density functions 
for Burgers turbulence,'' Phys. Rev. E \f.55., 6907 (1997).

\bb{13} S. N. Gurbatov, S. I. Simdyankin, E. Aurell, U. Frisch, and G. Toth, 
``On the decay of Burgers turbulence,'' J. Fluid Mech. \f.344., 339 (1997).

\bb{14} S. A. Boldyrev, private communication (1997).

\bb{15} Y. Kimura and R. H. Kraichnan, ``Statistics of an advected passive 
scalar,'' Phys. Fluids A \f.5., 2264 (1993), Sec. II.

\clbib

\end{multicols}

\eject

\parindent 0mm \parskip 3mm

FIGURE CAPTIONS

FIG. 1. Plot of $\log_{10}[\av{\xi^2}^{1/2} Q(\xi)]$ against 
$\xi/\av{\xi^2}^{1/2}$ for three sets of runs with forcing spectrum (B) and $R 
\sim 15$ (Run3, light dashes), 1200 (Run 4, heavy dashes) and 18000 (Run 5, 
solid line).

FIG. 2. The central region of $\xi_fQ(\xi)$ plotted against $\xi/\xi_f$ for 
the three runs with forcing spectrum (B), $R \sim 15$ (Run 3, light solid 
line), 1200 (Run 4, heavy solid line), 18000 (Run 5, heavy dashes), and a run 
with forcing spectrum (A) at $R \sim 15$ (Run 1, light dashes).

FIG. 3. Plot of $\log_{10}[\xi_f Q(\xi)]$ against $\log(|\xi|/\xi_f)$ for 
$\xi < 0$. $R \sim 15$ (Run 3, light dashes) $R \sim 1200$ (Run 4, heavy 
dashes) and $R \sim 18000$ (Run 5, solid line) are shown. The straight lines 
have slopes of $-3$ and $-1$.

FIG. 4. The central part of $\xi_fQ(\xi)$ plotted against $\xi/\xi_f$ for 
four cases: (a) the $R \sim 1200$ simulation (Run 4, heavy solid line); (b) 
the left-to-right numerical solution of \rf;4; in steady state with 
dissipation term set to zero (light solid line); (c) the left-to-right 
numerical solution with dissipation term taken as $0.45\xi_fQ(\xi)$ (heavy 
dashes); (d) the left-to-right solution with dissipation term taken as 
$0.8609\xi_fQ(\xi)/(1+\xi^2/xi_f^2)$ (light dashes).

FIG. 5. Plot of $3B\part(\ln Q)/\part(\xi^3)$ vs $\xi^3/B$ for a $R \sim 15$ 
simulation (Run 2, solid line). Dashed line is a mapping approximation.

FIG. 6. Plot of $3\part(\ln Q)/\part(\ln\xi^3)$ against $\xi^3/B$ (Run 2, 
solid line). Dashed line is a mapping approximation.

FIG. 7. Plot of $\nu H(\xi)/\xi_f^2$ against negative values of $\xi/\xi_f$ 
for a $R \sim 15$ simulation (Run 3, solid line). Also shown are the parabolas 
$(\xi/\xi_f)^2$ (light dashes) and $(\xi^2+\xi_C\xi)/\xi_f^2$ (heavy dashes).

FIG. 8. Plot of $\xi_frP(\ud,r)$ from the $R \sim 1200$ simulation (Run 4) 
against $\ud/r\xi_f$ (dashed lines) for values  of $r$ that increase by 
factors of $\sqrt{2}$ from $rk_f \approx .008$ to $rk_f \approx 0.72$. The 
curve that is highest on the left side is $rk_f \approx 0.72$. Also shown is 
$\xi_fQ(\xi)$ plotted against $\xi/\xi_f$ (solid line).

FIG. 9. $P(\ud,r)/\xi_fr$ plotted against $ud/u_{rms}$ for the $R \sim 1200$ 
run. The curves denote the $r$ values of Fig. 8.

FIG. 10. Plot of $\xi_frP(\ud,r)$ from the $R \sim 18000$ simulation (Run 5) 
against $\ud/r\xi_f$ (dashed lines) for values of $r$ that increase from $rk_f 
\approx 0.0016$ to $rk_f \approx 0.82$. The curve that is highest on the left 
side is $rk_f \approx 0.82$. Also shown is $\xi_fQ(\xi)$ plotted against 
$\xi/\xi_f$ (solid line).

FIG. 11. $P(\ud,r)/\xi_fr$ plotted against $\ud/u_{rms}$ for the $R \sim 
18000$ run. The curves denote the $r$ values of Fig. 10.

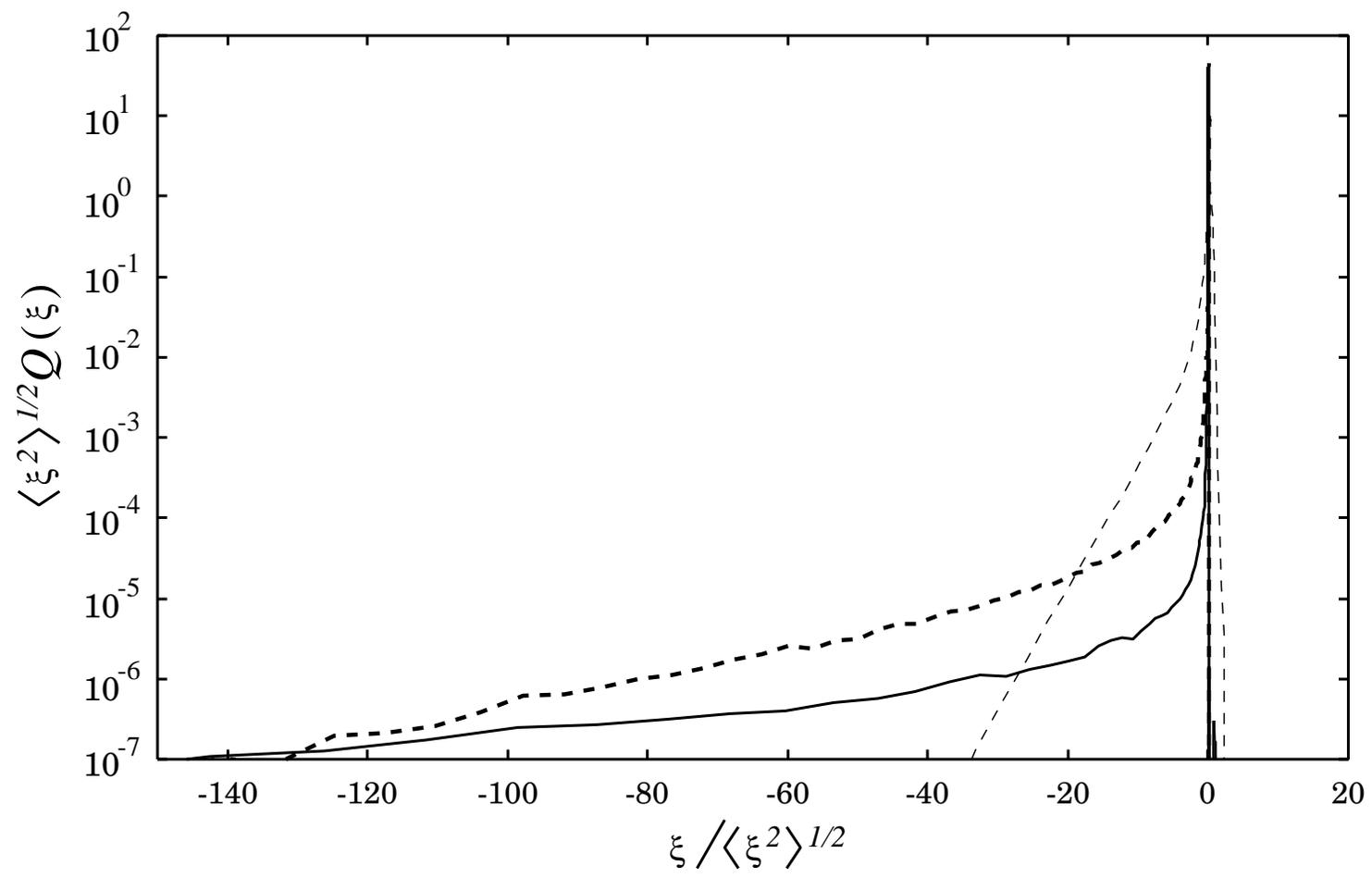
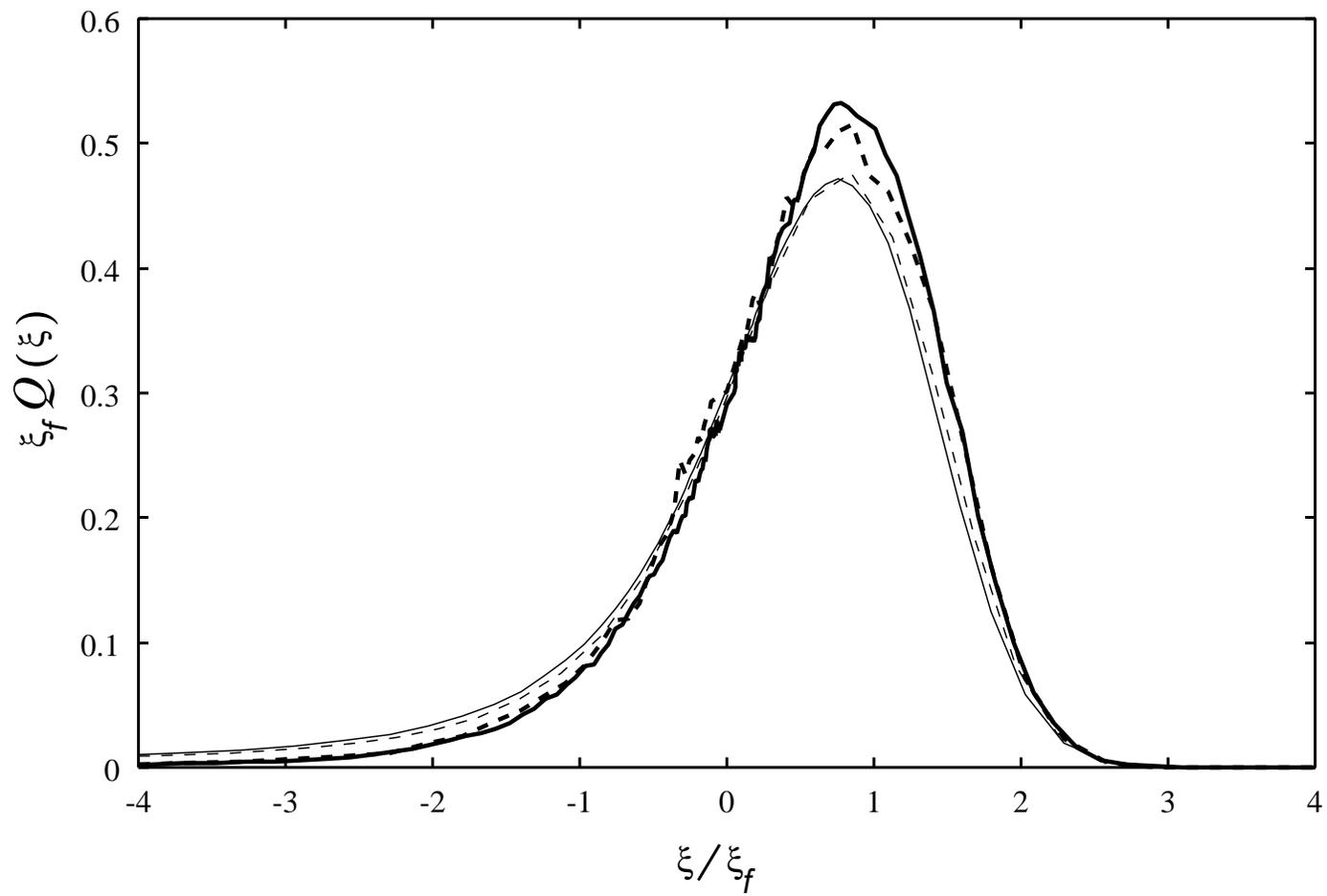
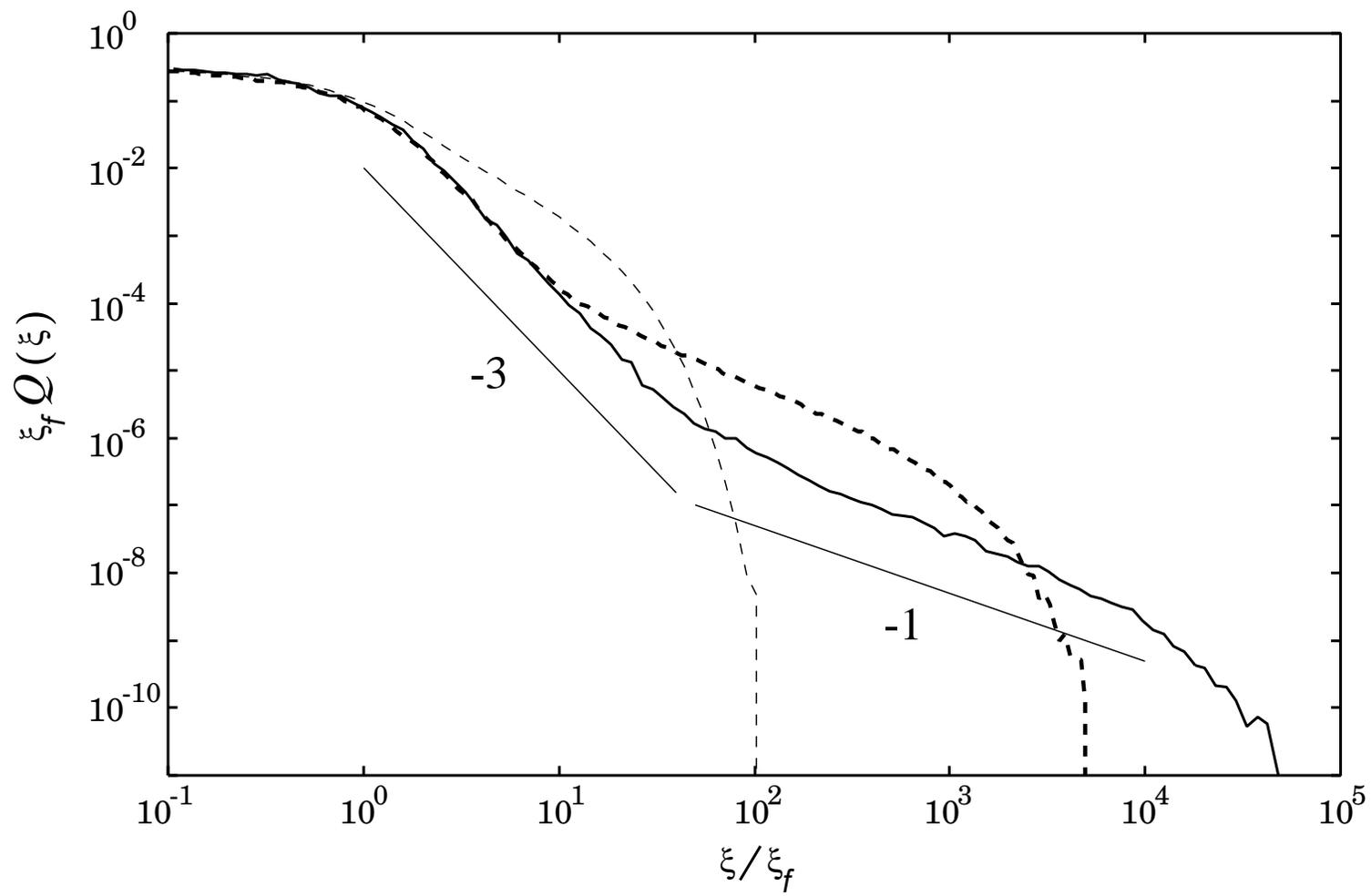
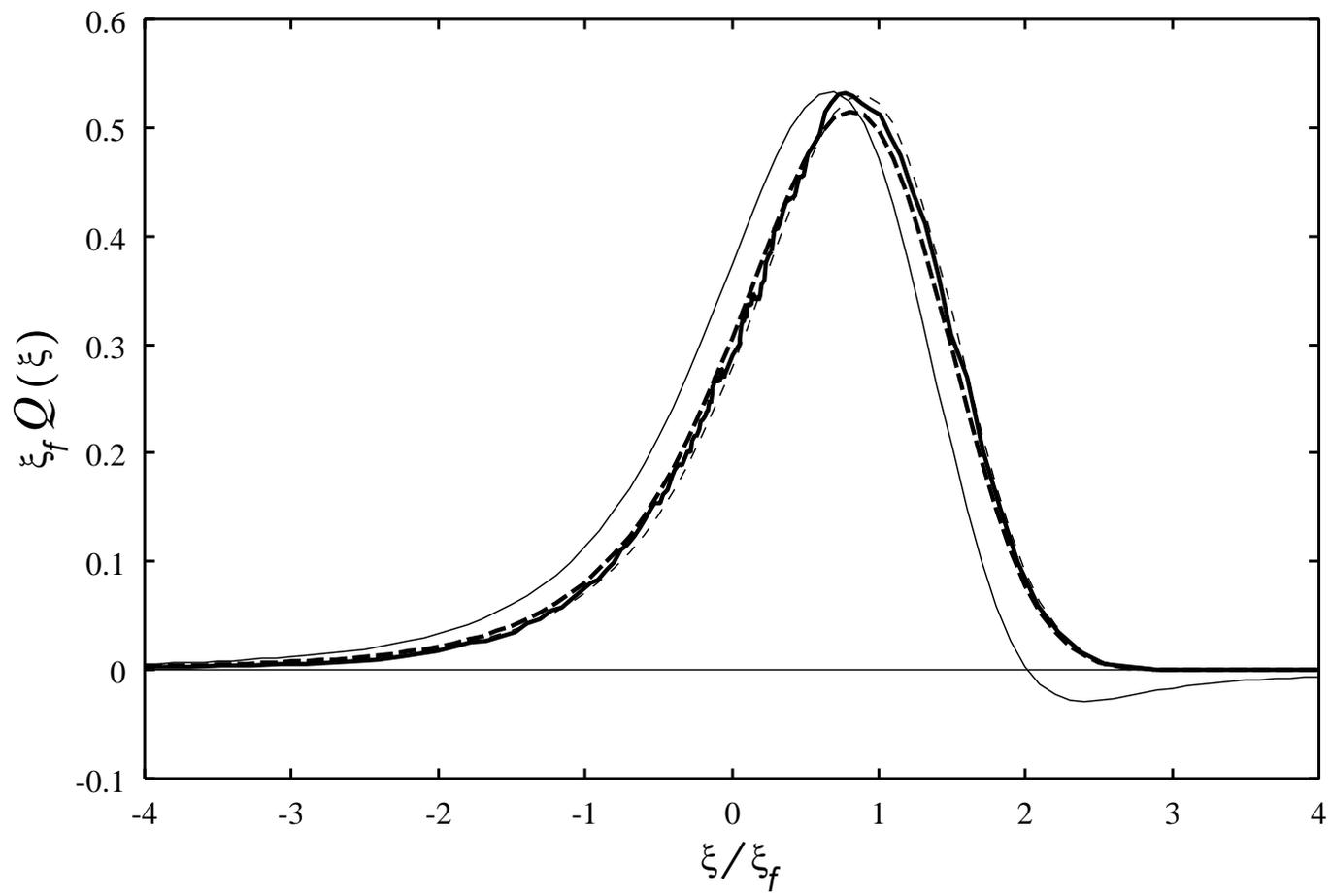
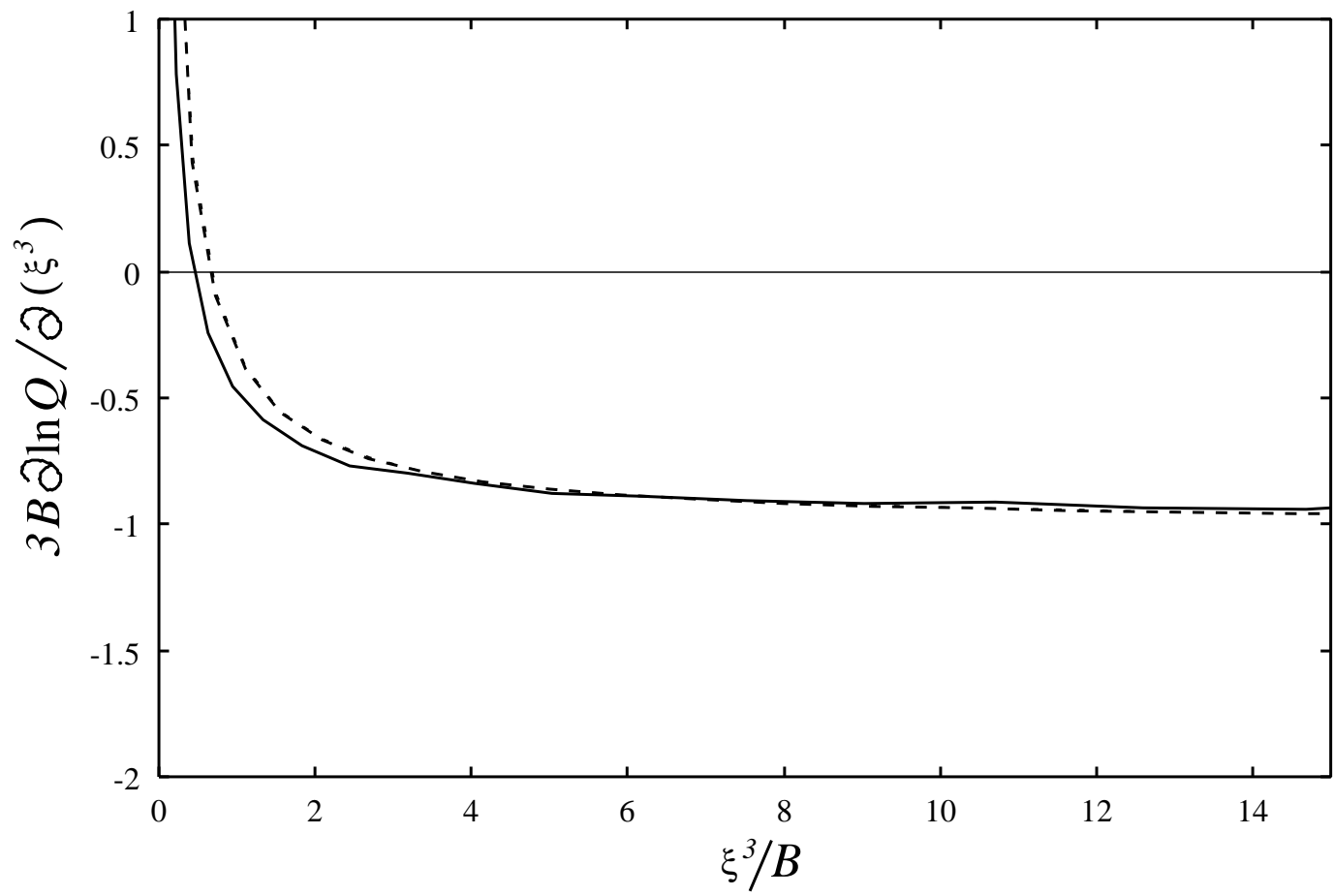
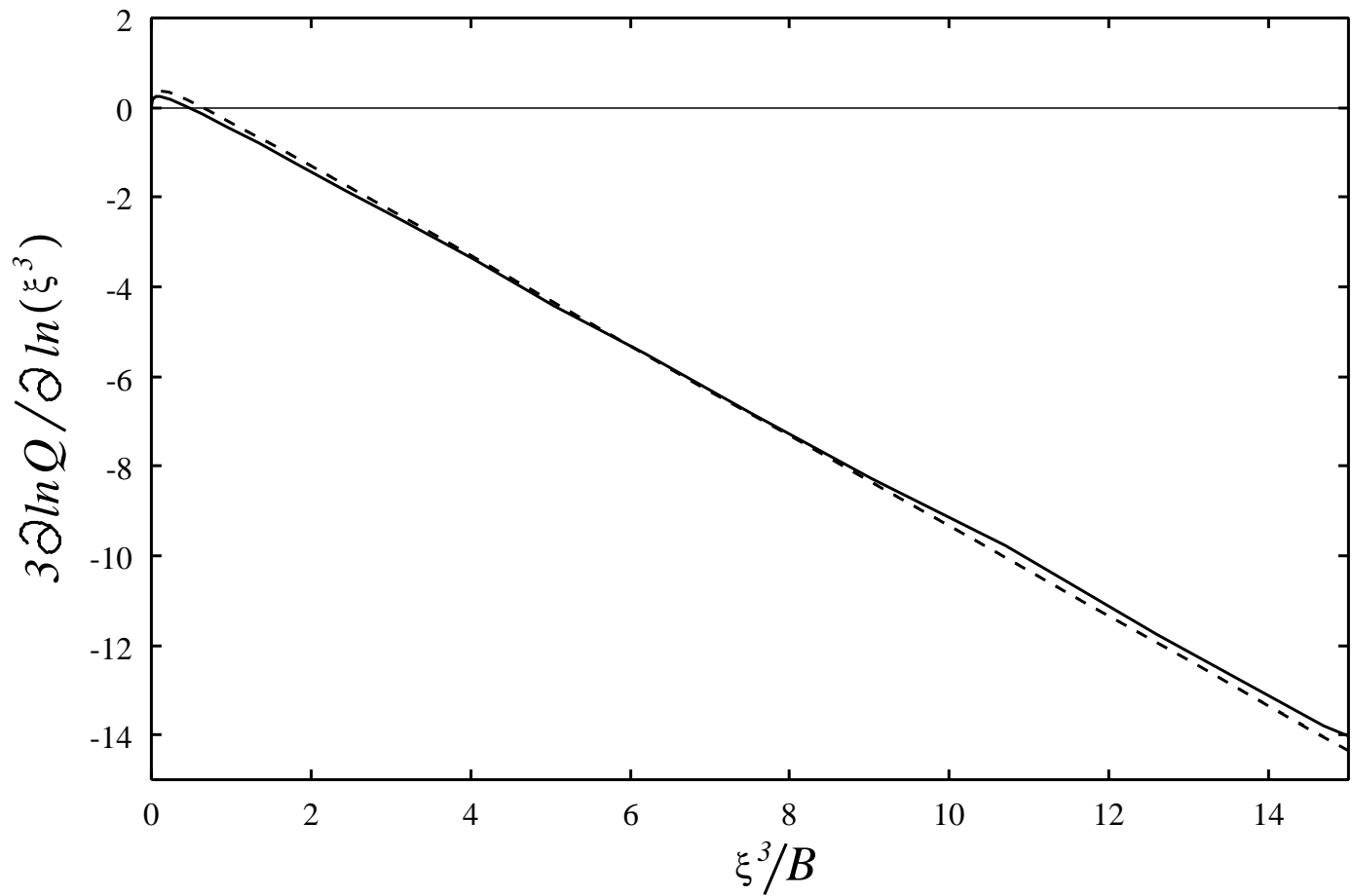
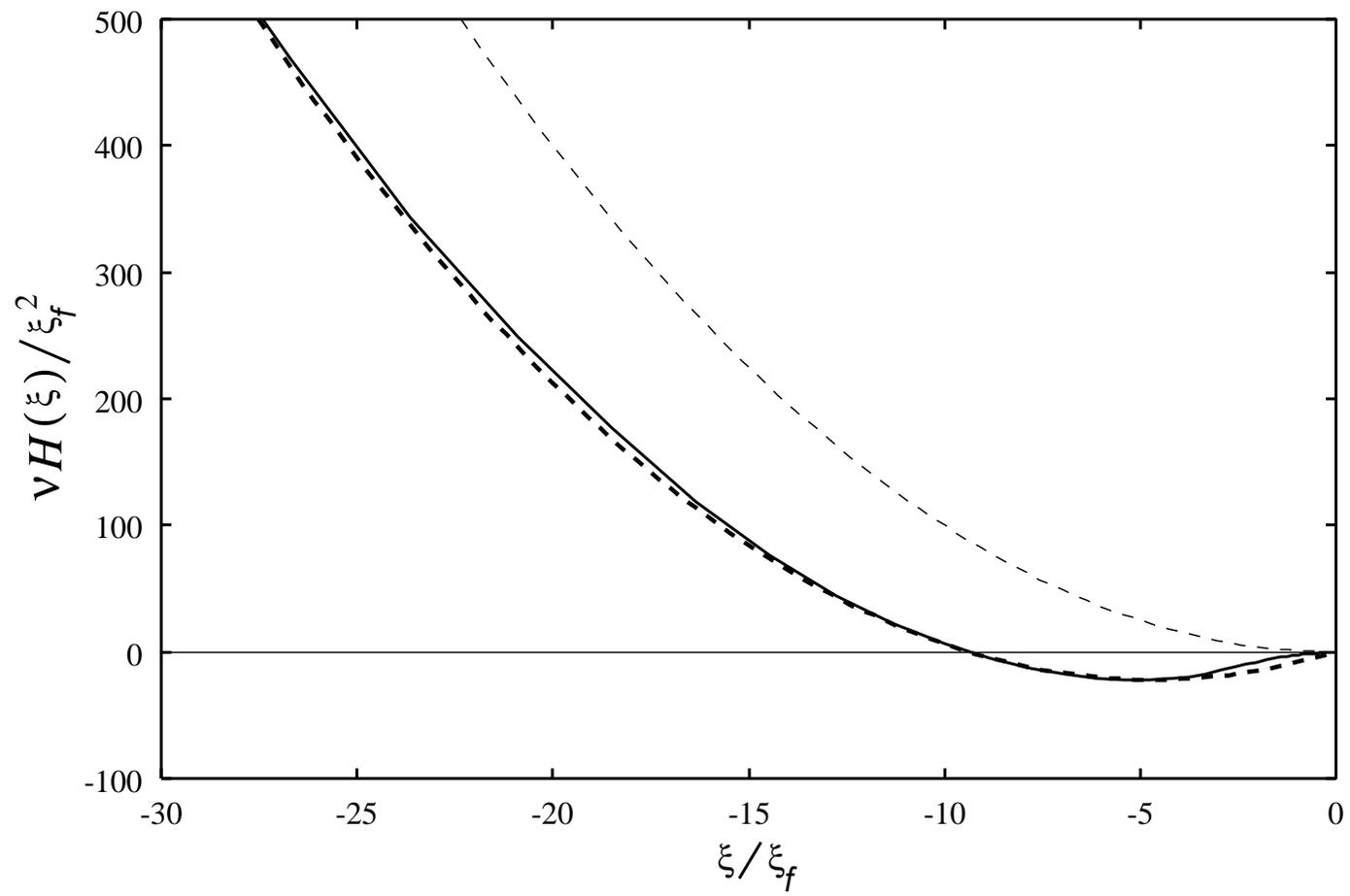
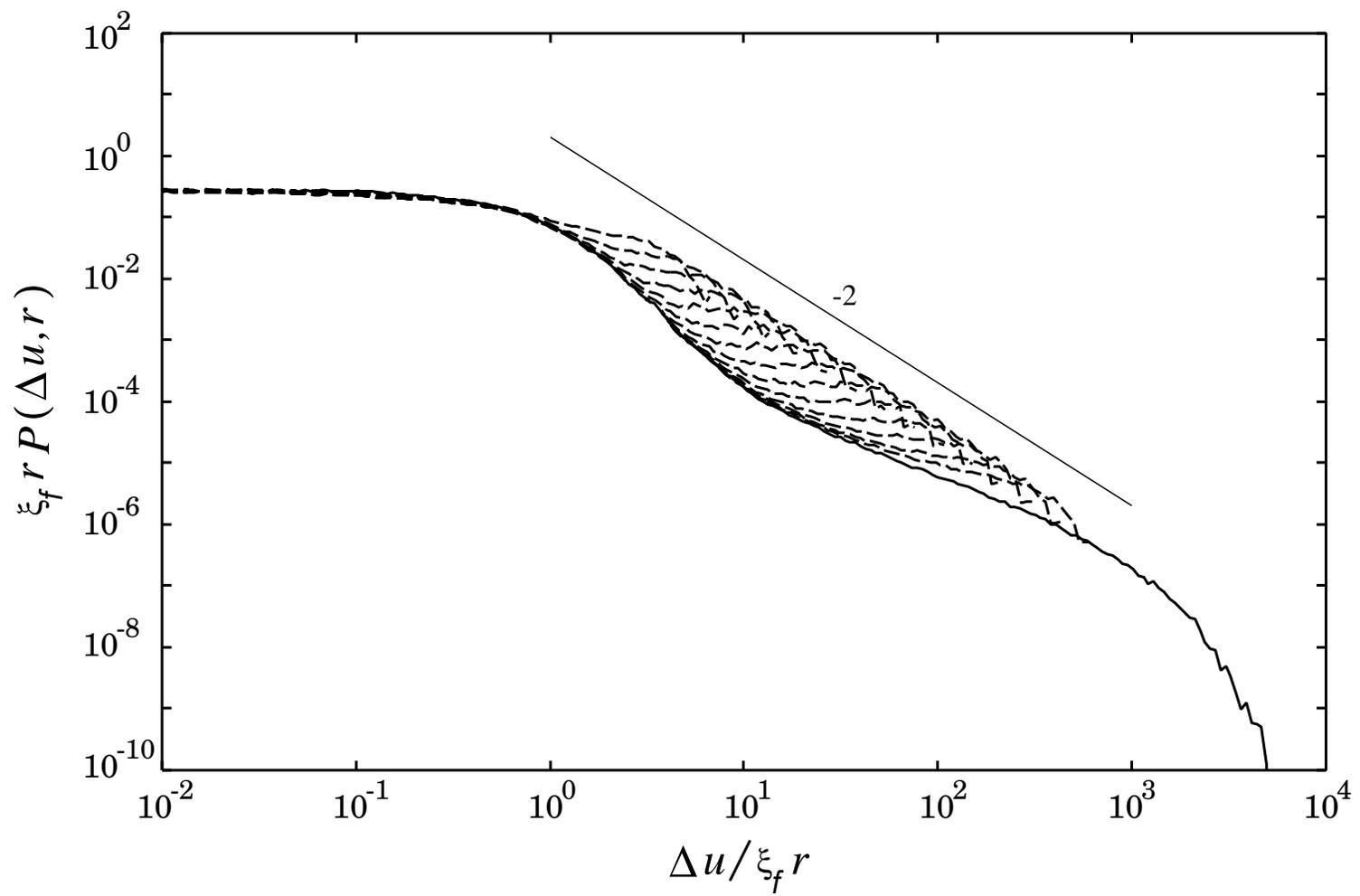
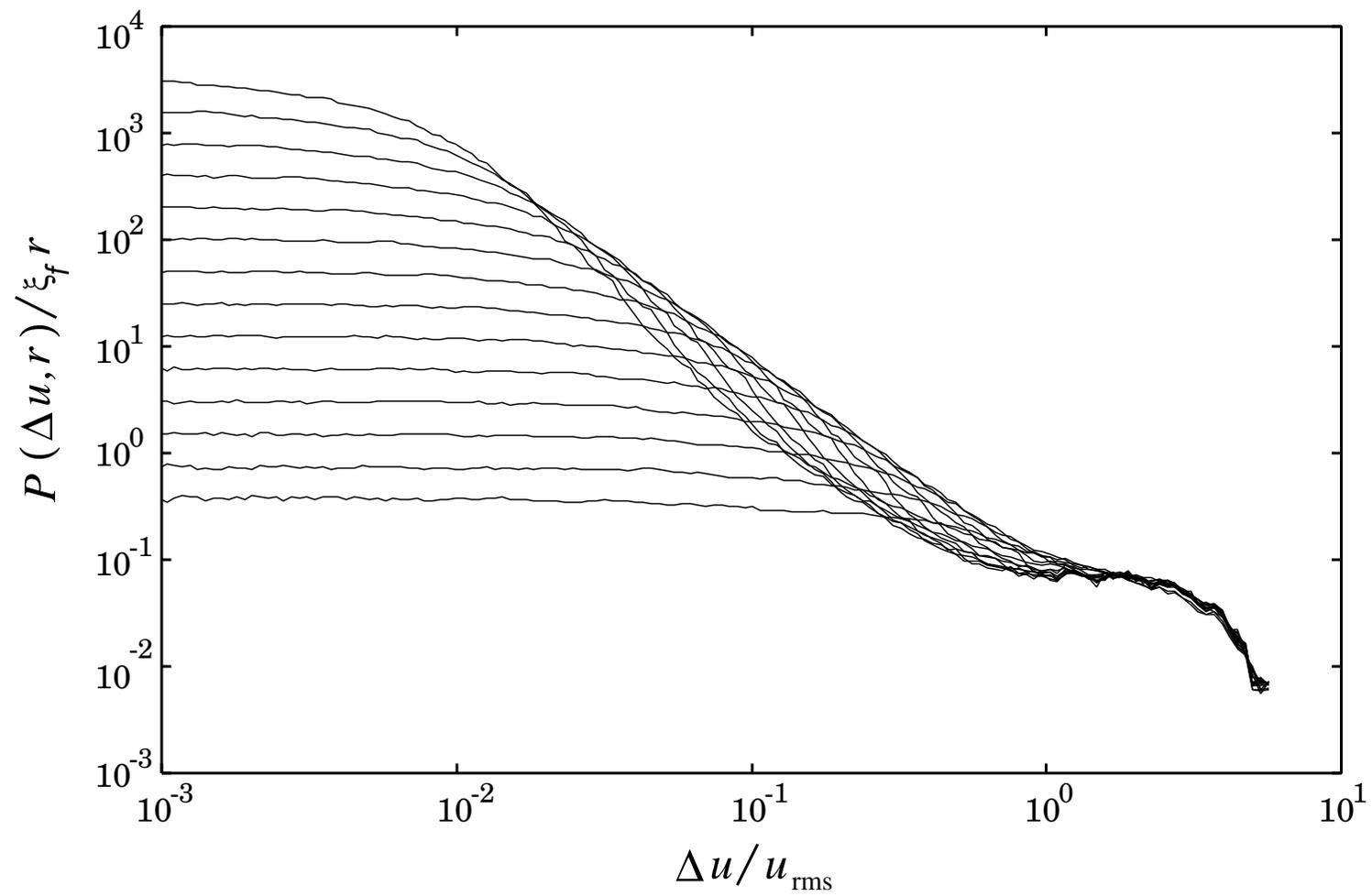
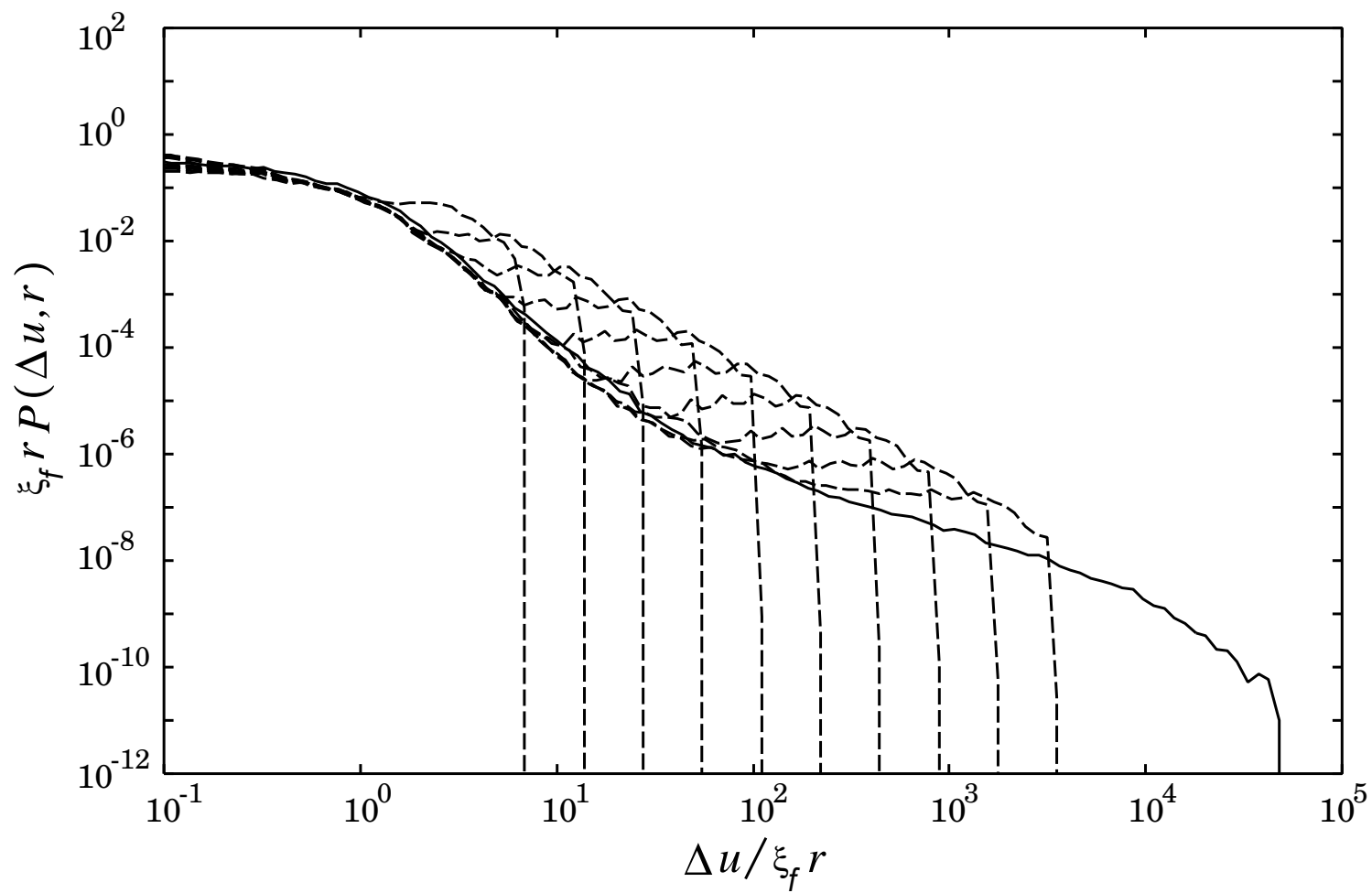
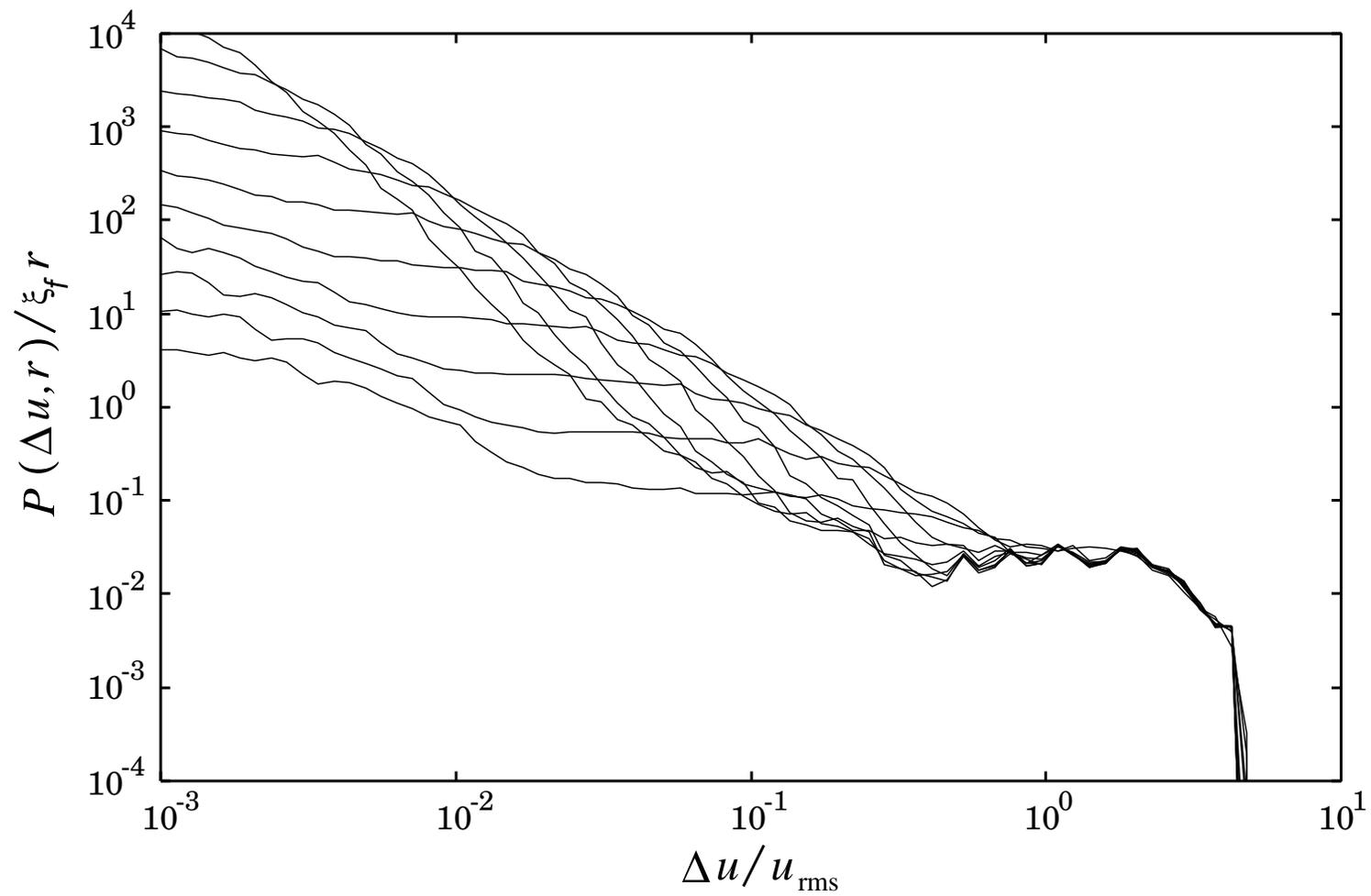
\end{document}